\documentclass[letter,bibyear]{aa} 

\usepackage{epsfig}
\usepackage{amsmath}
\usepackage{subfigure}
\usepackage{natbib}
\usepackage{multicol}
\usepackage{multirow}
\usepackage{color}
\usepackage{float}
\usepackage{longtable}
\usepackage{graphicx}
\usepackage{epstopdf}
\usepackage{booktabs}
\usepackage{txfonts}

\newcommand{\jms}{J.~Mol.~Spectrosc.}   
\newcommand{\jmst}{J.~Mol.~Struct.}   

\newcommand{\kms}{km s$^{-1}$}

\begin{document}

\title{Discovery of two metallic cyanoacetylides in IRC+10216: HMgCCCN and NaCCCN\thanks{Based on observations carried out with the Yebes 40m telescope (projects 19A010, 20A017, 20B014, 21A019). The 40m radio telescope at Yebes Observatory is operated by the Spanish Geographic Institute (IGN, Ministerio de Fomento).
}}

\author{
C.~Cabezas\inst{1},
J.~R.~Pardo\inst{1},
M.~Ag\'undez\inst{1},
B.~Tercero\inst{2,3},
N.~Marcelino\inst{2,3},
Y. Endo\inst{4},
P.~de~Vicente\inst{3},
M.~Gu\'elin\inst{5},
J.~Cernicharo\inst{1}
}

\institute{Departamento de Astrof\'isica Molecular, Instituto de F\'isica Fundamental (IFF-CSIC),
C/ Serrano 121, 28006 Madrid, Spain\\
\email carlos.cabezas@csic.es \& jose.cernicharo@csic.es
\and Centro de Desarrollos Tecnol\'ogicos, Observatorio de Yebes (IGN), 19141 Yebes, Guadalajara, Spain
\and Observatorio Astron\'omico Nacional (OAN, IGN), Madrid, Spain
\and Department of Applied Chemistry, Science Building II, National Yang Ming Chiao Tung University, 1001 Ta-Hsueh Rd., Hsinchu 300098, Taiwan
\and Institut de Radioastronomie Millim\'etrique, 300 rue de la Piscine, F-38406,
  Saint Martin d'H\`eres, France
}
\date{Received; accepted}

\abstract{ We report on the detection of a series of six lines in the ultra-deep Q-band integration toward IRC +10216 carried out with the Yebes 40m telescope, which are in harmonic relation with integer quantum numbers $J_u$ from 12 to 18. After a detailed analysis of all possible carriers, guided by high-level quantum chemical calculations, we conclude that the lines belong to HMgCCCN, named hydromagnesium cyanoacetylide. The rotational temperature and column density derived for HMgCCCN are 17.1\,$\pm$\,2.8\,K and (3.0\,$\pm$\,0.6)\,$\times$\,10$^{12}$ cm$^{-2}$, respectively. The observed abundance ratio between MgCCCN and HMgCCCN is $\sim$3. In addition, we report the discovery in space, also toward IRC\,+10216, of sodium cyanoacetylide, NaCCCN, for which accurate laboratory data are available. For this species we derive a rotational temperature of 13.5\,$\pm$\,1.7\,K and a column density of (1.2\,$\pm$\,0.2)\,$\times$\,10$^{11}$ cm$^{-2}$.}


\titlerunning{HMgC$_3$N and NaC$_3$N in IRC\,+10216}
\authorrunning{Cabezas et al.}

\maketitle

\section{Introduction}

Metal-containing molecules are almost exclusively found in space toward circumstellar envelopes around evolved stars, except for a few recent detections toward protostars \citep{Ginsburg2019,Tachibana2019}. Most of these molecules have been discovered toward  IRC\,+10216 the carbon-rich envelope of the star CW Leo. The first metal-bearing molecules detected in IRC\,+10216 were diatomic halides such as NaCl, KCl, AlCl, and AlF \citep{Cernicharo1987}. These species are formed in the hot inner parts of the envelope, close to the AGB star. The discovery and mapping of MgNC \citep{Kawaguchi1993,Guelin1993} revealed that metal-bearing molecules are also formed in the cool outer regions of the envelope. Other metal cyanides and isocyanides, such as NaCN \citep{Turner1994}, MgCN \citep{Ziurys1995}, AlNC \citep{Ziurys2002}, KCN \citep{Pulliam2010}, FeCN \citep{Zack2011}, HMgNC \citep{Cabezas2013}, and CaNC \citep{Cernicharo2019a}, were later on detected in the same source and are probably formed in the cool outer regions of the envelope. Indeed, metal atoms have been observed to survive in the gas phase in these regions \citep{Mauron2010}, allowing for the formation of these metal-bearing molecules, specially those containing magnesium. Indeed, a variety of Mg-containing carbon chains of increasing length, such as MgC$_2$, MgCCH, MgC$_4$H, MgC$_6$H, MgC$_3$N, and MgC$_5$N, have been recently detected \citep{Agundez2014,Cernicharo2019b,Pardo2021,Changala2022}.

In this letter we present the discovery in space of two new metal-bearing carbon chains toward IRC\,+10216. These molecules are HMgCCCN, a further member of the family of Mg-bearing molecules, and NaCCCN, the first long carbon chain of the family of Na-bearing molecules. These two molecules were detected thanks to a deep integration in the Q band (31.0-50.3 GHz) with the Yebes\,40m telescope. The identification of HMgCCCN is based on high-level quantum chemical calculations while that of NaCCCN relies on previous laboratory data \citep{Cabezas2019}. In addition, we provide an upper limit to the abundance of the aluminum cyanoacetylide, AlCCCN, in IRC\,+10216.

\section{Observations} \label{observations}

\begin{table*}
\small
\centering
\caption{Observed line parameters of HMgCCCN and NaCCCN toward IRC\,+10216.}
\label{line_parameters}
\begin{tabular}{cccccccc}
\hline
\hline
Transition&$\nu_{obs}$$^a$    &$\nu_{obs}-\nu_{cal}$$^b$&$\int$ $T_A^*$ dv $^c$ & $T_A^*$(horn)\,$^d$ & $T_A^*(center)$\,$^e$
& Notes\\
 $J_u$-$J_l$         & (MHz)             & (MHz)                   & (mK km s$^{-1}$)      & (mK)                & (mK)      &          & \\
\hline
HMgCCCN\\
\hline
  12-11& 31672.704$\pm$0.050& 0.0569& 39.84& 1.48& 1.31& {$^A$}\\
  13-12& 34311.875$\pm$0.050&-0.0304& 38.53& 1.45& 1.26& \\
  14-13& 36951.092$\pm$0.050&-0.0407& 42.28& 1.59& 1.39& \\
  15-14& 39590.327$\pm$0.020&       &      &     &     & {$^B$}\\
  16-15& 42229.489$\pm$0.050& 0.0040& 53.36& 2.06& 1.72& {$^C$}\\
  17-16& 44868.617$\pm$0.050& 0.0118& 50.25& 2.22& 1.48& \\
  18-17& 47507.686$\pm$0.050& 0.0010& 72.11& 3.06& 2.19& \\
  19-18& 50146.722$\pm$0.040&       &      &     &     & $\le$3.5, $^D$\\

\hline
NaCCCN\\
\hline
12-11&31705.51$\pm$0.20 & -0.0447  &13.25& 0.49& 0.44& \\
13-12&34347.56$\pm$0.30 &  0.0147  &11.35& 0.43& 0.37& {$^E$}\\
14-13&36989.38$\pm$0.20 & -0.1235  &12.17& 0.46& 0.40& \\
15-14&39631.36$\pm$0.20 & -0.0667  &11.35& 0.44& 0.37& \\
16-15&42273.31$\pm$0.20 & -0.0024  &10.64& 0.41& 0.34& \\
17-16&44915.22$\pm$0.40 &  0.0619  &17.38& 0.77& 0.51& {$^F$}\\
18-17&47556.89$\pm$0.40 & -0.0709  &17.67& 0.75& 0.54& \\
\hline
\hline
\end{tabular}
\tablefoot{\\
\tablefoottext{a}{Observed frequency assuming a $V_{LSR}$\,=\,$-$26.5 \kms.}
\tablefoottext{b}{Observed minus calculated frequencies in MHz.}
\tablefoottext{c}{Integrated line intensity in mK\,km\,s$^{-1}$. The total uncertainty is assumed to be dominated by the calibration uncertainties of 10\,\%.}
\tablefoottext{d}{Antenna temperature at the terminal velocity (horn) in mK.}
\tablefoottext{e}{Antenna temperature at line center in mK.}
\tablefoottext{A}{Line partially blended with a weak hyperfine component of CC$^{13}$CH. Frequency and line parameters can be correctly estimated adopting the observed intensities of the other hyperfine components of CC$^{13}$CH (see top panel of Fig.~\ref{fig_hmgcccn}).}
\tablefoottext{B}{Line fully blended with C$_3$N. Frequency and line parameters cannot be estimated. The quoted frequency corresponds to the predicted one.}
\tablefoottext{C}{Line partially blended with $c$-C$_3$H$_2$. Frequency and line parameters can still be estimated.}
\tablefoottext{D}{Line not detected. The upper limit in antenna temperature corresponds to 3$\sigma$. The quoted frequency corresponds to the
predicted one.}
\tablefoottext{E}{Line partially blended with an unknown feature. Frequency and line parameters can still be estimated.}
\tablefoottext{F}{Line blended with an unknown feature and with a line from HC$_9$N in the 3$\nu_{19}$ state. Frequency and line parameters are very uncertain.}
}
\end{table*}

New receivers, built within the Nanocosmos\footnote{ERC grant ERC-2013-Syg-610256-NANOCOSMOS.\\
https://nanocosmos.iff.csic.es/} project and installed at the Yebes\,40m radiotelescope, were used for the observations of IRC\,+10216 ($\alpha_{J2000}=9^{\rm h} 47^{\rm  m} 57.36^{\rm s}$ and $\delta_{J2000}=+13^\circ 16' 44.4''$) as part of the Nanocosmos survey of evolved stars \citep{Pardo2022}. Data from the QUIJOTE\footnote{\textbf{U}ltrasensitive \textbf{I}nspection \textbf{J}ourney to the \textbf{O}bscure \textbf{T}MC-1 \textbf{E}nvironment} line survey \citep{Cernicharo2021a,Cernicharo2022,Cernicharo2023} toward TMC-1 were also used to discriminate between possible carriers of the lines found in IRC\,+10216.

\begin{figure}
\centering
\includegraphics[width=0.89\columnwidth]{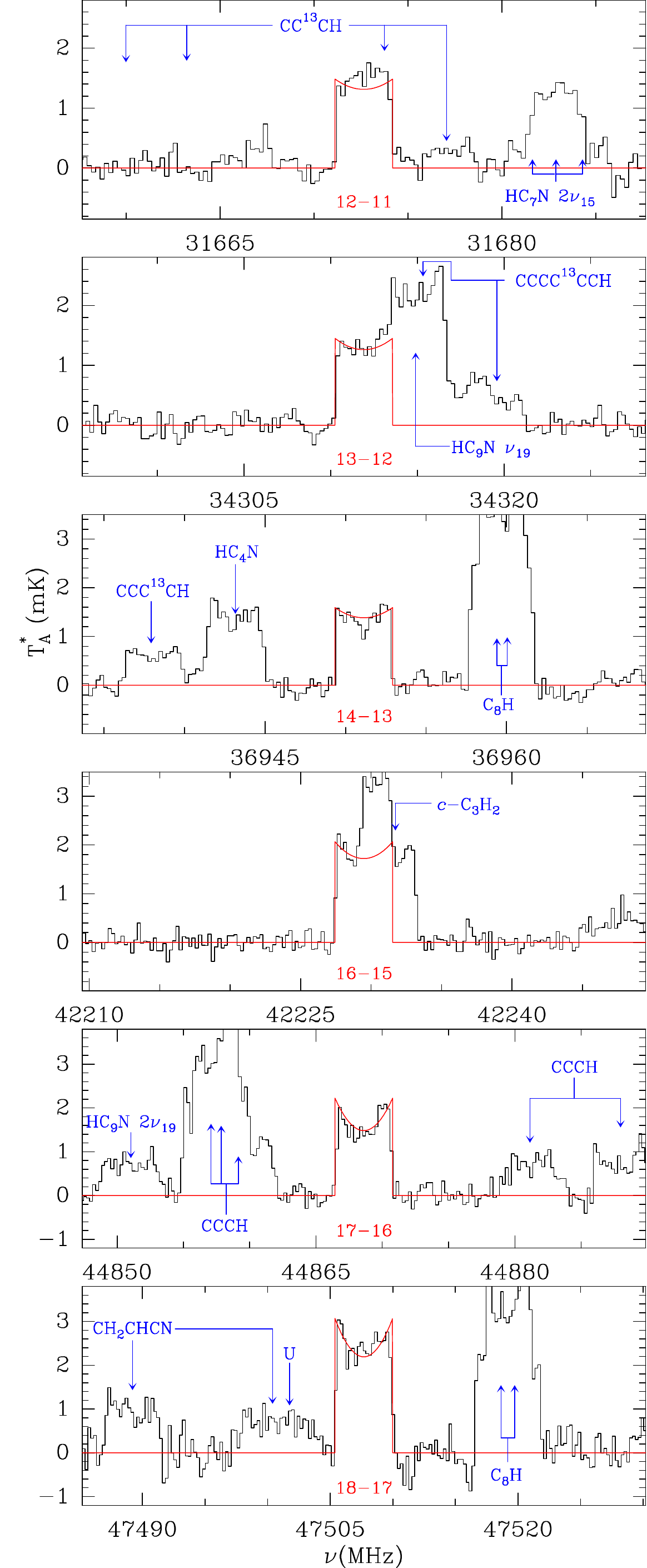}
\caption{Lines of HMgCCCN observed with the Yebes\,40m telescope toward IRC\,+10216. Line parameters are given in Table~\ref{line_parameters}.
The abscissa corresponds to the rest frequency assuming a local standard of rest velocity of $-$26.5 km s$^{-1}$. The ordinate is the antenna temperature corrected for atmospheric and telescope losses in mK. The red lines show the fitted line profiles adopting an expanding terminal velocity of 14.5 \kms \citep{Cernicharo2000}.}
\label{fig_hmgcccn}
\end{figure}

\begin{figure}
\centering
\includegraphics[width=0.88\columnwidth]{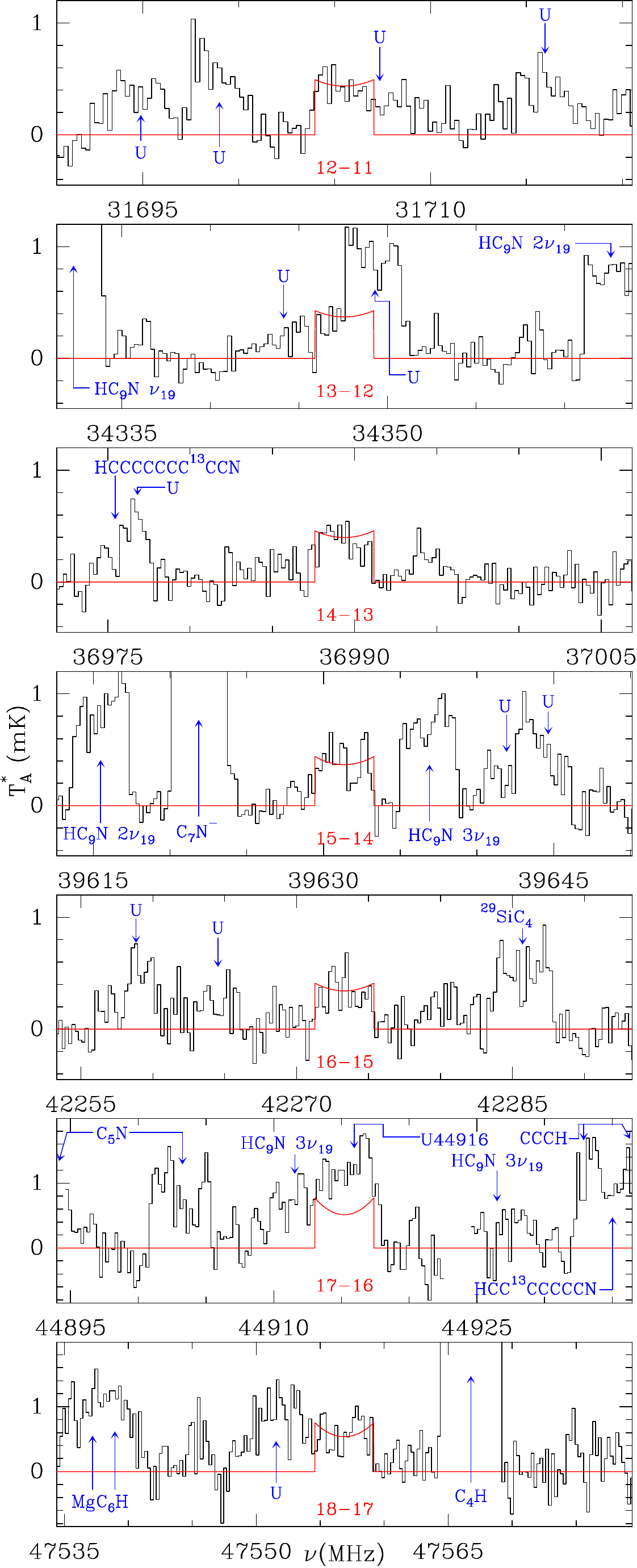}
\caption{Same as Fig.~\ref{fig_hmgcccn} but for NaCCCN.}
\label{fig_nacccn}
\end{figure}

A detailed description of the telescope, receivers, and backends can be found in \citet{Tercero2021}. Briefly, the receiver consists of two cold high electron mobility transistor amplifiers covering the 31.0-50.3 GHz band with horizontal and vertical polarizations. Receiver temperatures range from 16\,K at 31 GHz to 25\,K at 50 GHz. The backends are $2\times8\times2.5$ GHz fast Fourier transform spectrometers with a spectral resolution of 38.15 kHz, providing the whole coverage of the Q band in both linear polarizations. The data were smoothed to 228.9 kHz (six channels), which corresponds to a velocity resolution of 1.7\,km\,s$^{-1}$ at 40 GHz. This spectral resolution is good enough to resolve the broad U-shaped lines of IRC\,+10216 which exhibit a full velocity width of 29 km\,s$^{-1}$ \citep{Cernicharo2000}. The intensity scale used in this work, antenna temperature ($T_A^*$), was calibrated using two absorbers at different temperatures and the atmospheric transmission model ATM \citep{Cernicharo1985, Pardo2001}. Calibration uncertainties were adopted to be 10~\%. Additional uncertainties could arise from the line intensity fluctuation with time induced by the variation of the stellar infrared flux \citep{Cernicharo2014,Pardo2018}. The beam efficiency of the Yebes 40m telescope in the Q band is given as a function of frequency by $B_{\rm eff}$\,=\,0.797\,$\exp$[$-$($\nu$(GHz)/71.1)$^2$] and the forward telescope efficiency is 0.97. The emission size is assumed to have a radius of 15 arcsec, as observed for MgNC IRC+10216 \citep{Guelin1993}.
Pointing corrections were obtained by observing the SiO masers of R Leo. Pointing errors were always within 2-3$''$. All data were analyzed using the GILDAS package\footnote{\texttt{http://www.iram.fr/IRAMFR/GILDAS}}.

\section{Results} \label{results}

Our current data of IRC\,+10216 have never before reached a sensitivity ($\sigma$) as low as 0.15 mK at some frequencies in the $T_A^*$ scale for a spectral resolution of 229 kHz. This exceptional data quality has already been shown in recent publications presenting the detection of MgC$_5$N and MgC$_6$H \citep{Pardo2021}, and of C$_7$N$^-$ \citep{Cernicharo2023}. The data show a large number of lines coming from isotopologues and vibrationally excited states of abundant species that are identified using the catalogues MADEX \citep{Cernicharo2012}, CDMS \citep{Muller2005}, and JPL \citep{Pickett1998}.

In order to derive line parameters we use the SHELL method of the GILDAS package, which is well adapted for the line profiles observed in circumstellar envelopes. A variable window of 50-100 MHz around the line centre is used in this process. In all cases we fix the terminal expansion velocity of the envelope to 14.5 \kms \citep{Cernicharo2000}.

\subsection{Identification of HMgCCCN} \label{det_hmgcccn}

Among the unidentified lines that arise with high signal-to-noise ratio in the current Q-band spectrum of IRC\,+10216, six of them are in harmonic relation from $J_u$=12 to $J_u$=18. They were reported as unidentified features in the previous, much noisier, published version of the survey \citep{Pardo2022}. The lines are shown in Fig.~\ref{fig_hmgcccn} and the line parameters are provided in Table~\ref{line_parameters}. Most of the observed lines of this series are free of blending, except the $J$=15-14 which is fully blended with C$_3$N. Two of them ($J$=13-12 and $J$=16-15) have a marginal overlap with another feature but line fit is still possible. The lines do not show any evidence for spectroscopic broadening due to fine and/or hyperfine structure. They can be fitted to a standard line profile for an expanding envelope with a terminal expansion velocity of 14.5 \kms \citep{Cernicharo2000}. Hence, the carrier of the lines has a $^1\Sigma$ ground electronic state. The lines can be fitted to the standard Hamiltonian of a linear rotor providing the spectroscopic constants shown in Table \ref{spec_hmgcccn}. We name this carrier as B1320. We checked that values of $B$ divided by integers ranging from 2 to 6 do not match the observations as many lines would be missing.

The derived rotational constant is close to that of HC$_5$N (1331.333 MHz). All the isotopologues of this species can be discarded since they have already been reported in this source \citep{Pardo2022}. In addition, all possible isomers of HC$_5$N have rotational constants larger than that of B1320 \citep{Cernicharo2020a}. Other potential species could be the isotopologues of the anions C$_6$H$^-$ and C$_5$N$^-$, which have rotational constants $B$ of 1376.8630 MHz and 1388.8673 MHz, respectively. However, the intensities of C$_6$H$^-$ in our survey are 10-15 mK, and those of C$_5$N$^-$ are 4-6 mK. Hence, all the isotopologues of these species will have intensities well below the mK level, thus undetectable at the current level of sensitivity. Moreover, the distortion constants for all these possible candidates are around 30 Hz which is a factor of three lower than that of B1320. Distortion constants as large as the observed one are only provided by molecules containing a metal. Some species containing Si could have rotational constants around the observed value. For example SiC$_4$ has a rotational constant of 1510 MHz and exhibits lines of 10-15 mK in the Q-band. All its istopologues have been observed in the laboratory and none matches the rotational constant of B1320. Species such as SiCCCN \citep{Umeki2014} or CCCCS \citep{Hirahara1993} are open-shell species and cannot be considered as potential carriers.

AlCCCN and NaCCCN have been observed in the laboratory and have rotational and distortion constants similar to those found for B1320. NaCCCN has been detected in this work (see Section \ref{det_nacccn}), and all its possible isomers have rotational constants larger than that of B1320 \citep{Cabezas2019}. The intensities of the NaCCCN lines are 2-3 times lower than those of B1320. Although the rotational constants of the NaCCCN isotopologues could be close to that of B1320, the expected intensities are too low to match those observed for B1320. AlCCCN is not detected at the present level of sensitivity and its isomers, similar to NaCCCN, have too large rotational constants. Hence, we have to consider another metal such as Mg for which several species containing it have been detected in IRC\,+10216. Ab initio calculations by \citet{Cabezas2014,Cabezas2019} exclude also other possibilities such as molecules containing Ca.

Taking into account that MgCCCN and MgCCCCH are both detected in IRC\,+10216 with column densities of $\sim$10$^{13}$
cm$^{-2}$ \citep{Cernicharo2019b,Pardo2021}, we proceed in our search with high level ab initio calculations for their
hydrogenated counterparts HMgCCCN and HMgCCCCH using the same level of calculation as in previous studies
\citep{Cernicharo2019b}: CCSD(T)-F12 \citep{Knizia2009} with all electrons (valence and core) correlated and the
Dunning's correlation consistent basis sets with polarized core-valence correlation triple-$\zeta$ for explicitly
correlated calculations (cc-pCVTZ; \citealt{Hill2010}). All the calculations were carried out using the Molpro 2020.2
program \citep{Werner2020}. The calculated rotational constants for the two species are very similar: 1319.1 MHz and
1319.3 MHz for HMgCCCN and HMgCCCCH, respectively. However, the low dipole moment of 0.3\,D calculated for HMgCCCCH
excludes it as a possible carrier because a column density as high as $\sim$\,8\,$\times$\,10$^{14}$ cm$^{-2}$ would
be needed to reproduce the observed intensities. In contrast, the dipole moment calculated for HMgCCCN is 4.5\,D,
which makes it the best candidate for B1320. The centrifugal distortion constant $D$ calculated at the same level
of theory for HMgCCCN is 62.1 Hz, a bit different from the one derived in our fit. However, the hydromagnesium
derivatives are more floppy molecules, due to the HMgC bending mode, than the non-hydrogenated analogues and the
prediction of this parameter is less reliable \citep{Cernicharo2019b,Pardo2021}. This is also seen for HMgNC,
whose $D$ value is 2.94 kHz, while the one calculated using the mentioned level of calculation is 2.3 kHz. We
used this experimental/theoretical ratio to scale the theoretical value of $D$ for HMgCCCN with the result of
79.0 Hz, which is closer to the one derived from our astronomical data considering 3$\sigma$ uncertainty of the
experimental value. It must be said that we obtained a reliable fit also keeping $D$ constant fixed to 79.0 Hz.
From the observed line intensities of HMgCCCN we derive a rotational temperature of 17.1\,$\pm$\,2.8\,K and a column density of (3.0\,$\pm$\,0.6)\,$\times$\,10$^{12}$ cm$^{-2}$. The derived rotational temperature agrees well with the value derived for the cold MgC$_3$N component, T$_{rot}$=15$\pm$2\,K \citep{Cernicharo2019b}. However, the lines of MgC$_3$N in the 3mm domain, which involve high energy rotational levels, require an additional gas component with a rotational temperature of 34$\pm$6\,K.

\begin{table}
\small
\caption{Spectroscopic parameters of HMgCCCN}
\label{spec_hmgcccn}
\centering
\begin{tabular}{lccc}
\hline
\hline
 Parameter   &   Astronomical     & Theoretical$^a$\\
\hline
$B$ /MHz &    1319.7222$\pm$0.0021     &     1319.1 \\
$D$ /Hz&     99.2$\pm$4.1            &     79.0$^b$   \\
$rms$ /kHz &     38.7                    &            \\
$\mu$ / D   &                        &     4.5    \\
\hline
\hline
\end{tabular}
\tablefoot{\\
\tablefoottext{a}{Calculations from this work (see text).}
\tablefoottext{b}{Value corrected using the experimental/theoretical $D$ ratio of HMgNC (see text).}}
\end{table}
\normalsize

\begin{table}
\small
\caption{Spectroscopic parameters of NaCCCN.}
\label{NaCCCN_constants}
\centering
\begin{tabular}{lcc}
\hline \hline
\multicolumn{1}{c}{Parameter}  & \multicolumn{1}{c}{Lab + IRC\,+10216$^a$} & \multicolumn{1}{c}{Laboratory\,$^b$} \\
\hline
$B$        / MHz   &  ~ 1321.094464(27)$^c$   &  ~ 1321.094334(28) \\
$D$       /  Hz   &  ~     104.83(56)   &  ~     101.1(7) \\
$eQq$(Na) / MHz   &  ~    -7.0497$^d$     &  ~   -7.0497(7)  \\
$eQq$(N)  / MHz   &  ~    -3.8761$^d$    &  ~   -3.8761(5)  \\
$rms$$^e$     / kHz   &  ~     15.9          &  ~     0.9        \\
$N$$^f$               &  ~       129         &  ~      122 \\
\hline
\hline
\end{tabular}
\tablefoot{
\tablefoottext{a}{Merged molecular parameters from a fit to the laboratory and IRC\,+10216 frequencies.}\tablefoottext{b}{Molecular parameters derived by \cite{Cabezas2019}.}\tablefoottext{c}{Numbers in parentheses are 1$\sigma$ uncertainties in units of the last digit.} \tablefoottext{d}{Fixed to the value determined from the laboratory data.} \tablefoottext{e}{Standard deviation of the fit.}  \tablefoottext{f}{Number of transitions included in the fit.}
}\\
\end{table}

\subsection{Detection of NaCCCN and non detection of AlCCCN} \label{det_nacccn}

The rotational constants of AlCCCN and NaCCCN are available from laboratory experiments \citep{Cabezas2014,Cabezas2019}. For NaCCCN seven lines were found in our IRC+10216 Q-Band data with frequencies in very good agreement with those predicted from the laboratory measurements. They are shown in Fig. \ref{fig_nacccn} and their line parameters are given in Table \ref{line_parameters}. Using the IRC+10216 frequencies and those measured in the laboratory \citep{Cabezas2019}, we derive a new set of molecular constants for NaCCCN, using SPFIT \citep{Pickett1991}, which are given in Table \ref{NaCCCN_constants}.

The calculated dipole moment of NaCCCN is 12.9\,D, using the same level of theory as for HMgCCCN. The rotational temperature and column density derived for NaCCCN are 13.5\,$\pm$\,1.7\,K and (1.2\,$\pm$\,0.2)\,$\times$\,10$^{11}$ cm$^{-2}$, respectively.

In spite of the high accuracy of the frequency predictions for AlCCCN in the Q band (better than 25 kHz), we failed to detect any of its lines at the predicted frequencies. Assuming an expanding velocity of 14.5\,km\,s$^{-1}$ \citep{Cernicharo2000}, we derive a 3$\sigma$ upper limit to its column density of 8\,$\times$\,10$^{11}$ cm$^{-2}$.

\section{Discussion} \label{discussion}

It is interesting to compare the abundance derived for HMgCCCN with those derived for other Mg-bearing cyanides detected in IRC\,+10216. The column densities reported in IRC\,+10216 for these species are $N$(MgNC)\,=\,1.3\,$\times$\,10$^{13}$ cm$^{-2}$ \citep{Kawaguchi1993,Cabezas2013}, $N$(HMgNC)\,=\,6\,$\times$\,10$^{11}$ cm$^{-2}$ \citep{Cabezas2013}, $N$(MgCN)\,=\,7.4\,$\times$\,10$^{11}$ cm$^{-2}$ \citep{Ziurys1995,Cabezas2013}, $N$(MgCCCN)\,=\,9.3\,$\times$\,10$^{12}$ cm$^{-2}$ \citep{Cernicharo2019b}, and $N$(MgC$_5$N)\,=\,4.7\,$\times$\,10$^{12}$ \citep{Pardo2021}. Therefore, HMgCCCN is just three times less abundant than MgCCCN, in contrast with the smaller hydromagnesium analogue HMgNC, which is 20 times less abundant than MgNC.

\begin{figure}
\centering
\includegraphics[width=\columnwidth,angle=0]{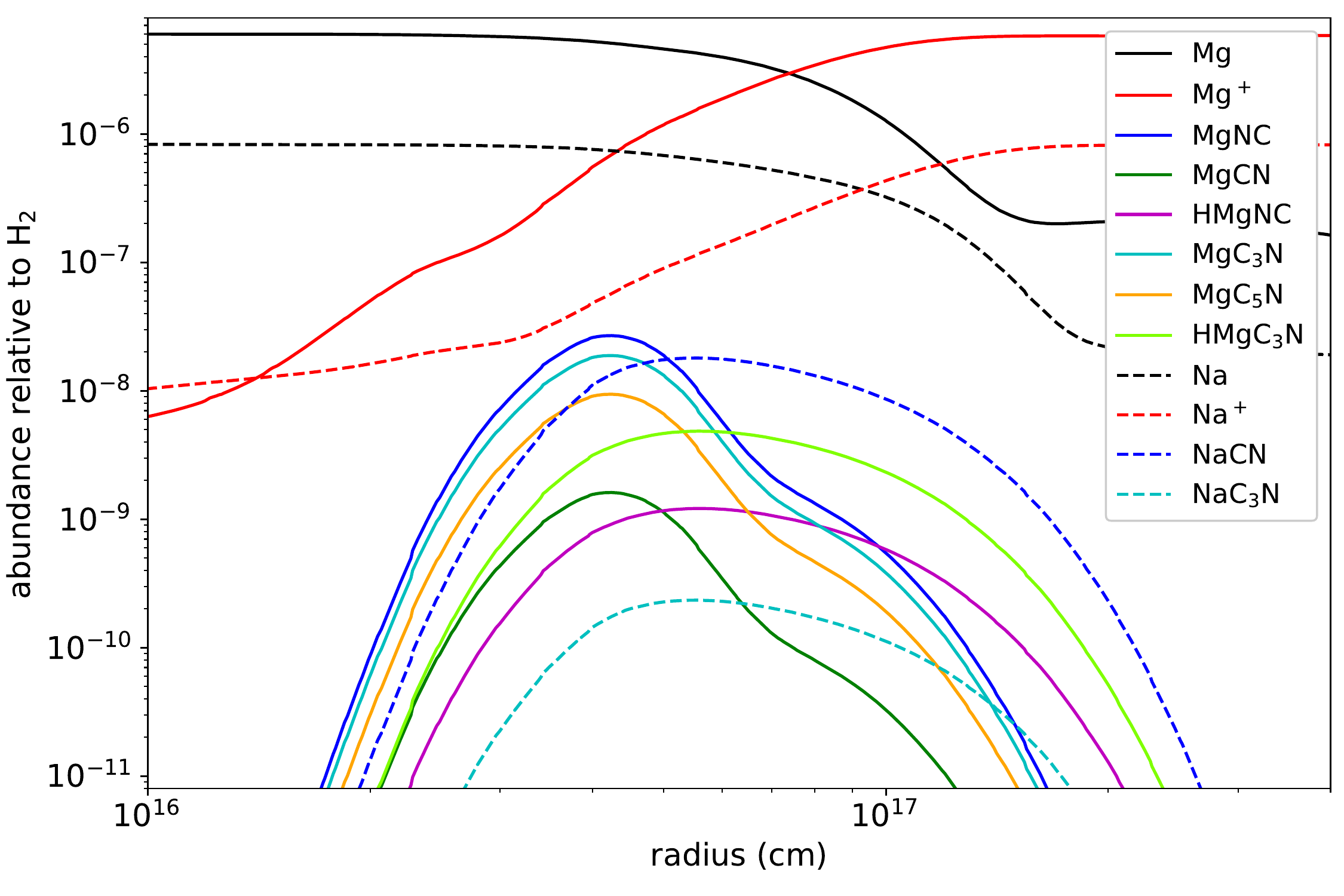}
\caption{Abundances calculated with the chemical model for Mg- and Na-bearing cyanides in IRC\,+10216. The initial abundance of Mg and the branching ratios of the dissociative recombination of metal-bearing cation complexes are chosen to reproduce the observed values. For example, the observed column density of MgNC is 1.3\,$\times$\,10$^{13}$ cm$^{-2}$ \citep{Cabezas2013} while the calculated value, evaluated as twice the radial column density, is 1.1\,$\times$\,10$^{13}$ cm$^{-2}$. Closed shell species, such as HMgNC, HMgC$_3$N, NaCN, and NaC$_3$N, extend further than open shell ones because we assume that they do not react with H atoms and electrons.}
\label{fig:model}
\end{figure}

In the case of sodium the number of Na-bearing cyanides detected in IRC\,+10216 is much smaller than for magnesium. The only such species detected in IRC\,+10216, apart from NaCCCN, is NaCN, which was first detected in IRC\,+10216 by \cite{Turner1994} and later on mapped by \cite{Guelin1997}. Unlike most metal-bearing cyanides detected in IRC\,+10216, which are most likely formed in the outer layers, NaCN is formed in the inner regions. The studies by \cite{Agundez2012} and \cite{Quintana-Lacaci2017} constrain the abundance of NaCN to (3-9)\,$\times$\,10$^{-9}$ relative to H$_2$, which implies that it is significantly more abundant than NaCCCN, as discussed below.

To shed light on the formation mechanism of HMgCCCN and NaCCCN in IRC\,+10216 we expanded the chemical model presented in \cite{Pardo2021}. The chemical scheme, based on the work of \cite{Petrie1996} and \cite{Dunbar2002}, starts with the injection of neutral metal atoms into the expanding wind. These atoms are then ionized by the interstellar radiation field and the resulting ionized metal atoms associate radiatively with long neutral carbon chains to form cationic complexes which then recombine dissociatively with electrons to yield neutral fragments as products that are detected in IRC\,+10216. The formation scheme of HMgCCCN is:
\begin{equation}
\rm Mg^+ + HC_{2n+1}N \rightarrow MgHC_{2n+1}N^+ + h\nu,
\end{equation}
\begin{equation}
\rm MgHC_{2n+1}N^+ + e^- \rightarrow HMgC_3N + C_{2n-2},
\end{equation}
where $n$ = 2, 3, 4 and the second reaction yields also other Mg-bearing neutral fragments in addition to HMgCCCN. For Mg we adopted an initial abundance of 3\,$\times$\,10$^{-6}$ relative to H, which allows to reproduce the observed column densities of Mg-bearing molecules, while for Na we adopted an abundance of 4.2\,$\times$\,10$^{-7}$ relative to H, as measured in the outer layers of IRC\,+10216 by \cite{Mauron2010}. The radiative association rate coefficients for ionized metal atoms and carbon chains are taken from the calculations of \cite{Dunbar2002}. The other critical input data in the chemical model are the branching ratios yielding the different fragments upon dissociative recombination of the cationic complexes. These ratios are not known and are difficult to predict. We therefore tuned them to reproduce the relative abundances of metal-bearing molecules observed in IRC\,+10216.

The results from the chemical model are shown in Fig.~\ref{fig:model}. The calculated abundances of Mg-bearing cyanides are relatively high, in the range 10$^{-9}$-10$^{-8}$ relative to H$_2$. In the case of sodium, we assumed that the main channel in the dissociative recombination of NaHC$_{2n+1}$N$^+$ yields NaCN, while NaCCCN is formed with a branching ratio of $\sim$\,1\,\%. Under this assumption, the column density of NaCCCN agrees with the observed one and NaCN is predicted to have an abundance of $\sim$\,10$^{-8}$ relative to H$_2$, within the range of values derived by \cite{Agundez2012} and \cite{Quintana-Lacaci2017}. Therefore, the observed abundance of NaCCCN is consistent with NaCN being around 100 times more abundant than it and with Na being injected into the expanding envelope with the abundance constrained by the observations of \cite{Mauron2010}.

\section{Conclusions}

We reported the first identification in space of two new metal-bearing carbon chains, HMgCCCN and NaCCCN, toward IRC\,+10216, the carbon-rich circumstellar envelope of CW Leo. The detection of HMgCCCN adds to the long list of Mg-bearing molecules already known to be present in IRC\,+10216. On the other hand, the detection of NaCCCN implies that long carbon chains containing metals other than magnesium are also formed in IRC\,+10216, which opens the door for future detections of carbon chains containing metals such as Na, Al, K, Fe, or Ca.

\begin{acknowledgements}
We thank ERC for funding
through grant ERC-2013-Syg-610256-NANOCOSMOS. We also thank Ministerio de Ciencia e Innovaci\'on of Spain (MICIU) for funding support through projects PID2019-106110GB-I00, PID2019-107115GB-C21 / AEI / 10.13039/501100011033, and PID2019-106235GB-I00.

\end{acknowledgements}

\end{document}